\documentclass[11pt]{article}
\usepackage[T1]{fontenc}
\usepackage{geometry}
\usepackage{microtype}
\usepackage{graphicx}
\usepackage{hyperref}
\usepackage{amsmath}
\usepackage{cleveref}
\usepackage{authblk}

\usepackage{newtxtext}
\usepackage{newtxmath}

\newcommand{\ddc}{\text{ddc}}
\DeclareMathOperator{\ind}{\mathbb{I}}
\DeclareMathOperator{\corr}{corr}
\DeclareMathOperator{\E}{\mathbb{E}}

\usepackage[american]{babel}
\usepackage{csquotes}
\usepackage[style=apa,giveninits=true,maxcitenames=1,backend=biber]{biblatex}
\DeclareLanguageMapping{american}{american-apa}

\title{Big data, big problems: Responding to ``Are we there yet?''}
\date{\today}

\author{Alex Reinhart}
\author{Ryan Tibshirani}
\affil{Delphi Group, Carnegie Mellon University}

\begin{document}
\maketitle

\begin{abstract}
  \textcite{Bradley:2021}, as part of an analysis of the performance of
  large-but-biased surveys during the COVID-19 pandemic, argue that the data
  defect correlation provides a useful tool to quantify the effects of sampling
  bias on survey results. We examine their analyses of results from the COVID-19
  Trends and Impact Survey (CTIS) and show that, despite their claims, CTIS in
  fact performs well for its intended goals. Our examination reveals several
  limitations in the data defect correlation framework, including that it is
  only applicable for a single goal (population point estimation) and that it
  does not admit the possibility of measurement error. Through examples, we show
  that these limitations seriously affect the applicability of the framework for
  analyzing CTIS results. Through our own alternative analyses, we arrive at
  different conclusions, and we argue for a more expansive view of survey
  quality that accounts for the intended uses of the data and all sources of
  error, in line with the Total Survey Error framework that have been widely
  studied and implemented by survey methodologists.
\end{abstract}

\section{Introduction}

\textcite{Bradley:2021}, in their article ``Are We There Yet? Big Surveys
Significantly Overestimate COVID-19 Vaccination in the US'', use recent surveys
of COVID-19 vaccination hesitancy and uptake to establish what they call a
paradox: ``the two `big surveys' are far more confident, yet also far more
biased, than the smaller, more traditional Axios--Ipsos poll.'' Using the
framework established by \textcite{Meng:2018}, they examine the possible sources
of error in the surveys, such as sampling bias, sample size, and population
heterogeneity. They conclude that for two of the surveys, sampling biases ``can
almost completely wipe out the statistical information'' in the data, rendering
suspect all point estimates and even correlations calculated using the survey
data.\footnote{We refer specifically to version 2 of their arXiv preprint,
  available at \url{https://arxiv.org/abs/2106.05818v2}. Because their work is
  under revision, future versions may differ.}

We are the designers and operators of one of the surveys discussed by
\textcite{Bradley:2021}: the COVID-19 Trends and Impact Survey (CTIS), operated
in the United States by the Delphi Group at Carnegie Mellon University in
collaboration with Facebook. (The survey is called ``Delphi-Facebook'' by
\textcite{Bradley:2021}.) This survey invites a random sample of participants
from Facebook's Active User Base each day, amounting to about 250,000 responses
in the United States per week. The survey has operated continuously since April
6, 2020, collecting over 22 million responses during that time and allowing
long-term tracking of COVID trends. Its sampling and weighting methodology is
discussed in more detail by \textcite{Kreuter:2020}, \textcite{Barkay:2020}, and
\textcite{Salomon:2021}; further documentation and public aggregated datasets
are available at
\url{https://cmu-delphi.github.io/delphi-epidata/symptom-survey/}.

\textcite{Bradley:2021} correctly note that CTIS systematically overestimates
COVID-19 vaccination rates in the United States, and we agree that users of any
large dataset should be aware that size is no guarantee of accuracy. However, we
disagree with their broader conclusions for several reasons:

\begin{enumerate}
\item As CTIS's name suggests, its design goals are to facilitate rapid
  detection of trends---such as sudden increases in symptom rates that may
  indicate a COVID hotspot---and to do so at a fine geographic level; the large
  sample size is necessary for these goals. These goals can be met even with
  systematic biases, provided the biases do not change or only change slowly
  over time. Our results indicate the survey has been largely successful in this
  goal.

\item \textcite{Bradley:2021} correctly note that observed increases in the data
  defect correlation (\textit{ddc}), the measure of ``data quality'' in the
  framework described by \textcite{Meng:2018}, do not necessarily imply a change
  in the sampling mechanism or an increase in sampling bias. But this point is
  worth greater emphasis, as it undermines the argument that the \textit{ddc}
  can play a useful role in the evaluation of large surveys. In fact, analyses
  of CTIS survey data show that it predicts both COVID case rates and COVID
  vaccination rates much more accurately than a simple \textit{ddc} analysis
  would show. For important problems such as geographic resource allocation, its
  effective sample size is thousands or tens of thousands of responses per week.

\item The data defect correlation framework is limited when applied to surveys
  because it does not account for sources of survey error beyond sampling
  biases. Self-reports from survey respondents can be biased for many reasons,
  and the \textit{ddc} is no longer easily interpretable when any such biases
  are present. Though \textcite{Bradley:2021} argue the \textit{ddc} is related
  to the design effect in this setting, we believe this is incorrect, and that
  it is unclear how to interpret the \textit{ddc} or how to use it to improve
  survey quality. This problem may also affect the scenario analyses they
  present in Section 6.

\item More broadly, while we agree with \textcite{Bradley:2021} that big surveys
  are not a panacea, we believe that there is no one number that summarizes the
  suitability of any survey for any purpose. The data defect correlation can be
  useful in some cases, but it does not help judge the suitability of a survey
  for tracking trends or for other purposes. Anyone planning to use a dataset,
  large or small, must carefully consider its suitability for their specific
  research goals, which may differ from those explored in the framework
  established by \textcite{Meng:2018}.
\end{enumerate}
So while \textcite{Bradley:2021} are correct to urge caution in the use of large
but biased datasets, we believe it is important to provide additional context
about our research goals, and to provide more actionable insights for
researchers evaluating large survey datasets. We discuss these points in greater
detail in the sections that follow.

\section{Background}

\subsection{CTIS design goals}

Since the value of a dataset depends on the research goals for which it is used,
it may be helpful to discuss the original goals of the COVID-19 Trends and
Impact Survey (CTIS). The Delphi Group at Carnegie Mellon University had prior
experience with influenza forecasting, often using both standard public health
surveillance data (such as reports of influenza-like illness from outpatient
doctor's visits) and unconventional signals, such as volumes of symptom-related
Google search queries \parencite[Chapter 4]{Farrow:2016}. This approach had been
successful with influenza, and Delphi's forecasts were among the most accurate
in annual forecasting challenges conducted by the Centers for Disease Control
\parencite[Table 3]{Lutz:2019}.

The CTIS effort began in March 2020, as efforts to track the COVID pandemic
accelerated. Delphi approached Facebook about the possibility of using its
platform to host symptom surveys that could identify rises in symptoms that may
precede increases in confirmed cases, providing early warning of COVID case
increases. Facebook agreed, and the effort quickly expanded to include an
international version of the survey, hosted by the University of Maryland. The
instrument also gained additional items on social distancing, mental health,
comorbidities, and other topics of public health interest. It has operated
continuously since its launch, and aggregate data is released publicly every
day, typically within 1--3 days of survey responses being received.

The survey's emphasis on tracking and hotspot detection informed its design. To
detect COVID hotspots, it would need to sample responses continuously so changes
in symptom rates could be detected within days. To localize these hotspots to
cities or counties, and thus facilitate public health responses, it would need
large sample sizes to ensure power to track trends separately for each major
city or county. Hence the sampling design ensured a response volume of tens of
thousands of responses per day.

This response volume could not help the survey accurately estimate the
population prevalence of COVID-19, but that was never the intention. In the
spring of 2020, the rate of asymptomatic cases of COVID-19 was unknown; even if
it was known, COVID-19 can present with many different symptoms, most of which
could easily be caused by other illnesses or even seasonal allergies. A symptom
survey would suffer from many sources of measurement error. But if symptom rates
suddenly increased, that could be a sign that public health officials should be
prepared for a case increase---even if no point estimate of prevalence could be
made.

\begin{figure}
  \centering
  \includegraphics[width=0.9\textwidth]{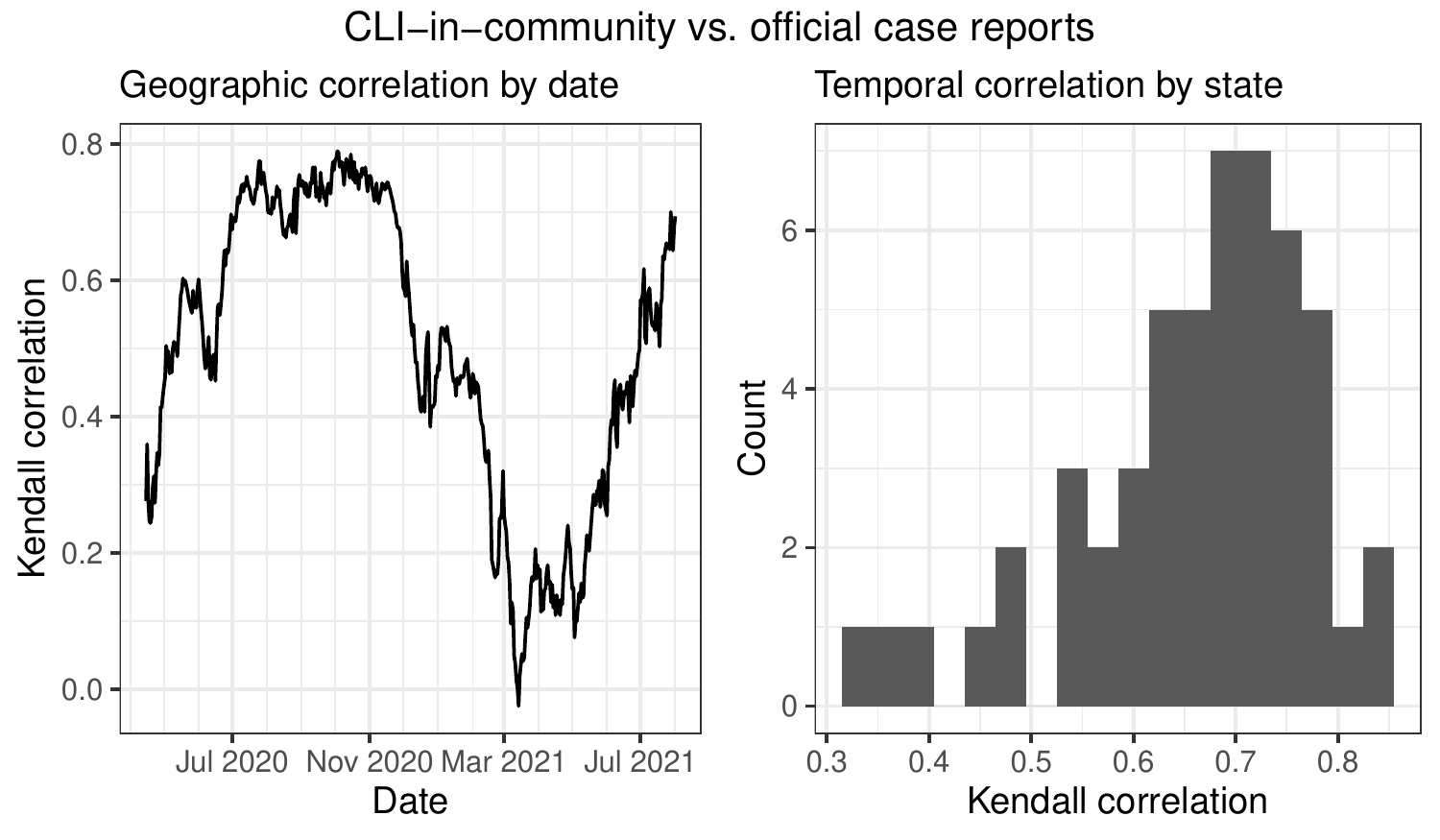}
  \caption{Correlation between the rate of CTIS respondents who know someone who
    is currently sick (``CLI-in-community'') and officially reported rates of
    COVID-19 cases (both as 7-day averages). \textit{Left:} On each date,
    correlation between CTIS estimates of CLI-in-community and state reports of
    new cases. \textit{Right:} For each state, correlation between the time
    series of CLI-in-community and of reported cases.}
  \label{cli-cases-correlation}
\end{figure}

This hypothesis was borne out by experience. Using the survey, we estimated
rates of COVID-like illness (CLI), as well as the percentage of respondents who
reported knowing someone in their local community who was currently sick (called
``CLI-in-community''). In \Cref{cli-cases-correlation}, we see the correlation
between CLI-in-community and official state reports of confirmed COVID cases.
For much of the pandemic, this correlation was quite strong, only dipping during
the winter of 2020 before recovering to again be quite high.\footnote{We
  speculate that this is at least partly due to the decline in reported cases in
  January 2021, resulting in there being less signal and more noise---i.e.,\
  there was less heterogeneity between states. We will return to this point in
  \Cref{bias-space}.} In fact, \textcite[Figs.~2 and 3]{Reinhart:2021} showed
that, for much of the pandemic, these CLI-in-community estimates correlated more
strongly with reported COVID cases than several other indicators, including some
based directly on medical claims from doctor's offices. \textcite{McDonald:2021}
further demonstrated that CLI estimates from the survey can be useful in
near-term COVID case forecasting, suggesting they contain information beyond
what is already available in public case reporting and hotspot detection. Survey
data is used to inform COVID-19 case and death forecasts submitted by several
teams to the CDC-sponsored COVID-19 Forecast Hub \parencite{Cramer:2021},
including forecasts by Delphi and by \textcite{Rodriguez:2021}.

In short, the sample size of CTIS was chosen to ensure it could meet its goals,
not to somehow compensate for potential weaknesses of its Facebook-based
sampling mechanism. Experience has shown that CTIS is useful for these goals. As
the pandemic progressed, data from CTIS also began to be used for numerous other
goals. For example, \textcite{Bilinski_2021} studied the public reaction to
increasing infection rates; \textcite{Sudre:2021jj} explored the strong
association between self-reported anosmia and test positivity (while showing
consistency between CTIS results and data collected through other means); and
\textcite{Lessler:2021cu} examined associations between in-person schooling,
mitigation measures, and COVID-19 test positivity. Nonetheless, we agree that
analysts using the data must be aware that it is best used for specific goals;
as \textcite{Kreuter:2020} wrote when introducing the survey,
\begin{quote}
  In order to minimize the impact of various sources of error, we recommend
  analysts focus on temporal variation. Although estimates for a single point
  in time may be affected by many error sources (stemming from both sample
  selection and measurement procedures) these errors are likely to remain
  constant over relatively short periods of time, thereby producing unbiased
  estimates of change over time (Kohler, Kreuter, \& Stuart, 2019).
\end{quote}
Our focus on changes over time and on spatial comparisons, as in
\Cref{cli-cases-correlation}, would be valid if the survey's sampling bias
indeed remains relatively constant over time and space. We will examine this in
\Cref{bias-space,bias-increasing}, after first reviewing the data defect
framework used by \textcite{Bradley:2021}.

\subsection{The data defect correlation framework}
\label{ddc-framework}

It will be helpful to review the error decomposition originally introduced by
\textcite{Meng:2018} and used by \textcite{Bradley:2021} to analyze the errors
in each survey's estimates.

Consider a population of individuals \(i = 1, \dots, N\). Suppose there is some
variable of interest \(Y_i\) that could be observed for each member of the
population; its population average is denoted \(\bar Y_N\). Now suppose we take
a random sample of size \(n\) from this population, and use the indicator
variable \(R_i \in \{0, 1\}\) to denote whether person \(i\) was included in the
random sample. Once we take the sample, we calculate the sample mean, which can
be written as
\[
  \bar Y_n = \frac{\sum_{i = 1}^N R_i Y_i}{\sum_{i=1}^N R_i}.
\]
Now let \(\hat \rho_{R,Y} = \corr_i(R_i, Y_i)\) be the correlation between
sampling and the random variable in the finite population, let \(f = n/N\), and
let \(\sigma_Y\) be the population standard deviation of \(Y_i\). Then
\textcite{Meng:2018} showed that one can decompose the estimation error of the
sample mean as follows:
\begin{equation}\label{error-decomposition}
  \bar Y_n - \bar Y_N = \underbrace{\hat \rho_{R,Y}}_\text{data quality} \times
  \underbrace{\sqrt{\frac{1 - f}{f}}}_\text{data quantity} \times
  \underbrace{\sigma_Y.}_\text{problem difficulty}
\end{equation}
The data quality term is the data defect correlation; ideally this correlation
is nearly 0. By rearranging \cref{error-decomposition}, we obtain
\begin{equation}\label{ddc-estimator}
  \hat \rho_{R,Y} = \frac{\bar Y_n - \bar Y_N}{\sqrt{\sigma^2_Y (1 - f)/f}}.
\end{equation}
If the population quantity \(\bar Y_N\) is known, this can be used to estimate
the data defect correlation. The \textit{ddc} results presented by
\textcite{Bradley:2021} use Centers for Disease Control data on COVID-19 vaccine
uptake as the ground truth, and plug in estimates \(\bar Y_n\) from various
surveys.

\section{CTIS estimates across states}
\label{bias-space}

\textcite{Bradley:2021} question whether a large but biased survey can
facilitate accurate geographic comparisons, even at the state level. Their
Figure 1 shows a weak correlation between CTIS estimates of vaccine uptake on
March 27, 2021 and official CDC data on state-level vaccine uptake, and this
serves as supporting evidence for their conclusion that (page 21):
\begin{quote}
  Selection bias tells us that respondents are not exchangeable with
  non-respondents, and hence it may impact all studies of that dataset to
  varying degrees. This includes study of associations -- both Delphi-Facebook
  and Census Household Pulse significantly overestimate the slope of vaccine
  uptake over time relative to that of the CDC benchmark (Fig. 2) -- as well as
  ranking -- the Census Household Pulse and Delphi-Facebook rankings are more
  correlated with each other ([Kendall] rank correlation: 0.49), than either
  ranking is with that of the CDC (0.21 and 0.26, respectively), as indicated in
  Fig. 1.
\end{quote}
In other words, because of the survey biases that presumably vary across space,
even studies of associations, such as state-level associations with factors that
might explain vaccine uptake, are suspect.

\begin{figure}
  \centering
  \includegraphics[width=0.8\textwidth]{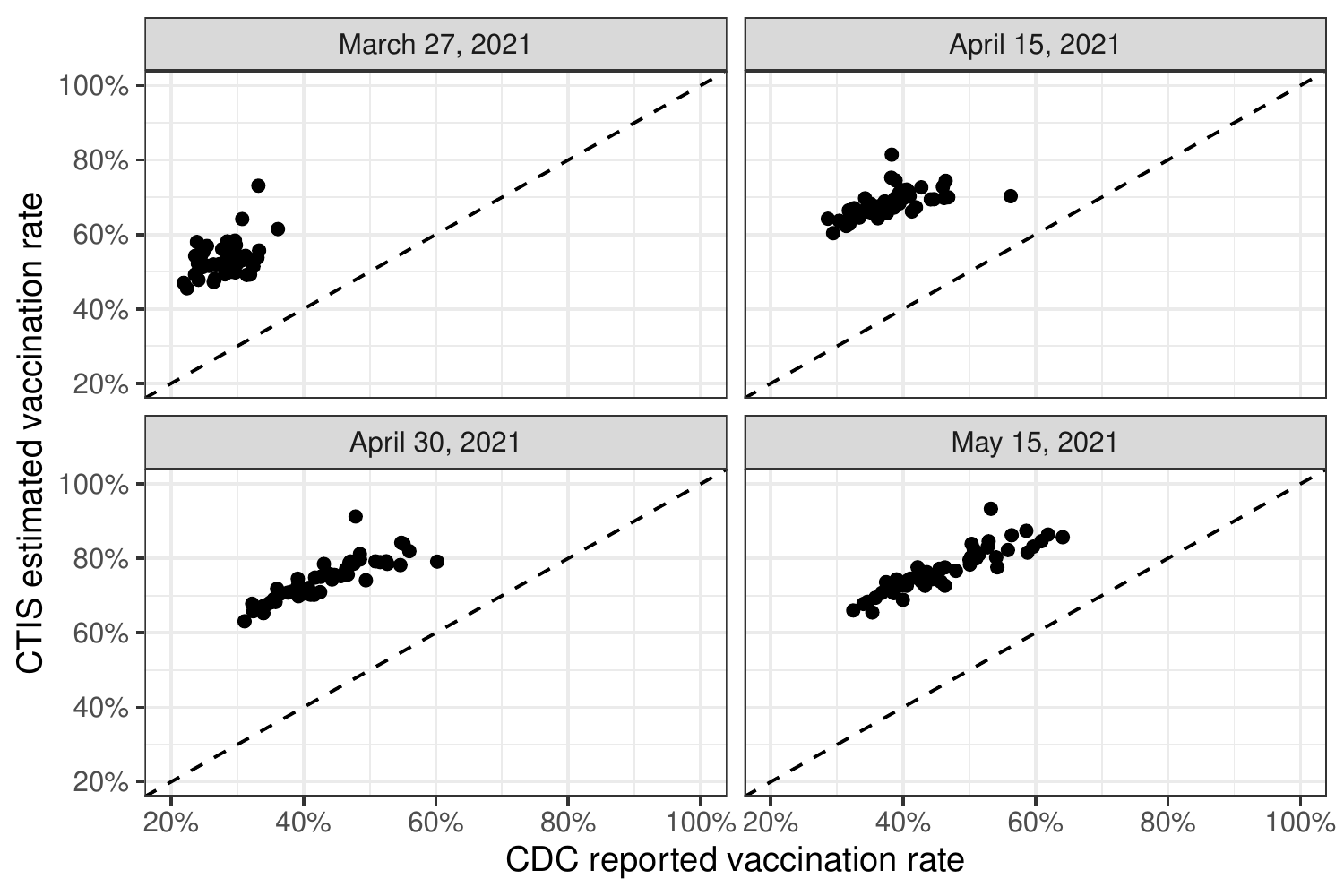}
  \caption{CDC-reported vaccination rates compared to CTIS estimates for each
    state. March 27th corresponds to Figure 1 of \textcite{Bradley:2021}, while
    the later dates show the rapid uptake of vaccines---and the increasing
    correlation between the survey and CDC data.}
  \label{cdc-compare-time}
\end{figure}

But these correlation results are based on vaccine uptake data as of March 27,
2021, several weeks before all adults in the United States became eligible to
receive vaccines. Because only seniors, healthcare workers, and certain other
high-risk and essential groups were eligible to receive vaccines, true
vaccination rates were quite low. In early April, all adults became eligible to
receive a vaccination, and vaccination rates rose rapidly over the next six
weeks. In \Cref{cdc-compare-time}, we see that the relationship between CTIS
estimates and CDC data firmed up substantially over this time.

\Cref{cdc-compare-time} hence raises two questions:
\begin{enumerate}
\item How did CTIS perform over time? The plots suggest it began to correlate
  well with CDC data as states began to diverge, but can we quantify this?
\item Is the lack of correlation in March due to spatial bias or due to the
  difficulty of the problem?
\end{enumerate}
We will examine each in turn.

\subsection{Comparison with CDC data over time}

\begin{figure}
  \centering
  \includegraphics[width=0.9\textwidth]{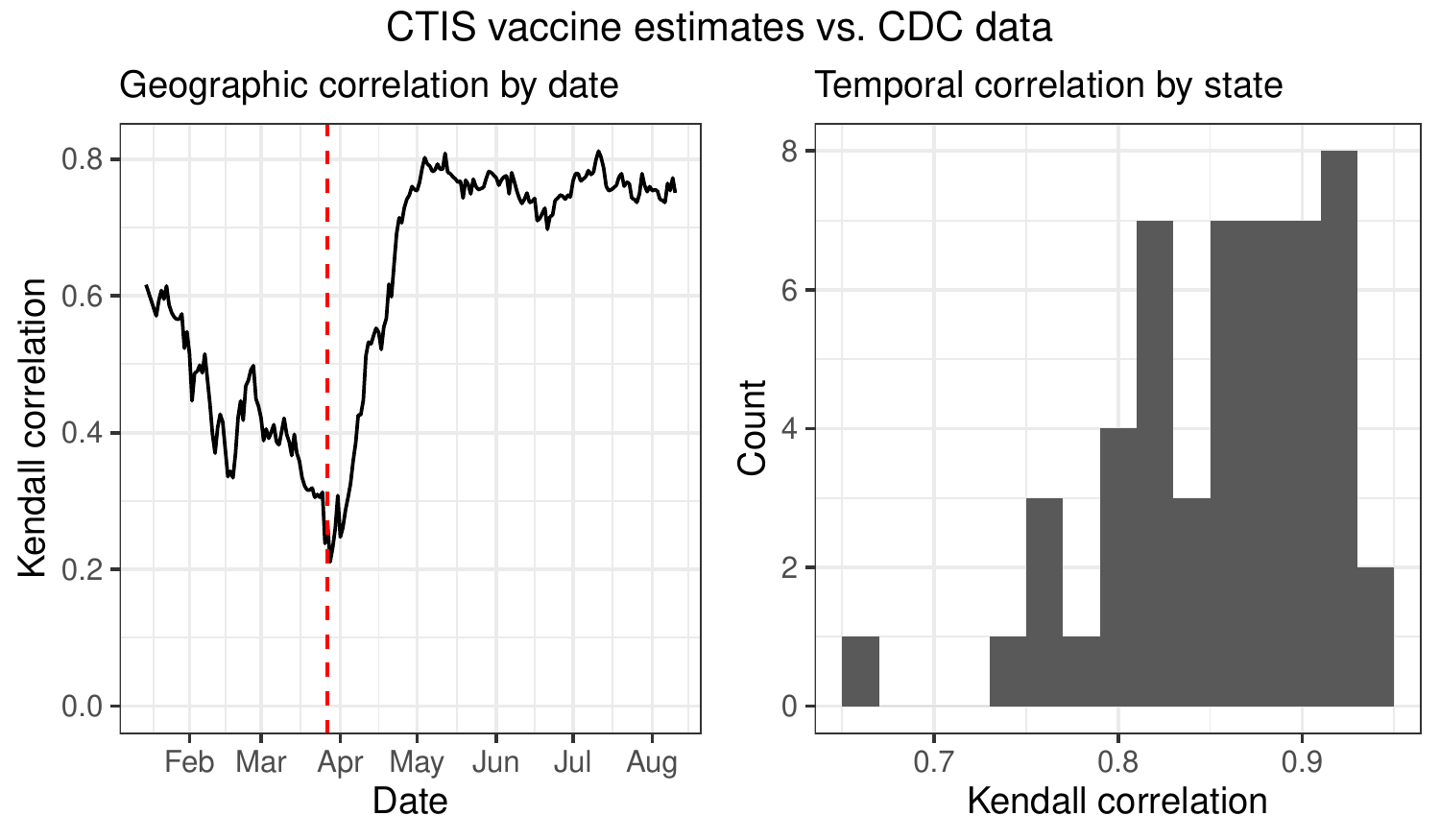}
  \caption{\textit{Left:} On each date, correlation between CTIS state-level
    estimates of vaccination rate and CDC data. March 27 is indicated with a red
    line. \textit{Right:} For each state, correlation between CDC and CTIS
    vaccination rate time series. Each time series shows high correlation, most
    over 0.8, indicating that CTIS tracks trends well. Wyoming shows the lowest
    Kendall correlation, 0.66.}
  \label{cdc-compare-correlation}
\end{figure}

We can make comparison in \Cref{cdc-compare-time} more concrete by considering
the Kendall correlation between CTIS estimates and CDC data over time, as the
vaccination campaign progressed. (We chose Kendall correlation to match the
analyses performed by \textcite{Bradley:2021}.) The left panel of
\Cref{cdc-compare-correlation} shows this correlation, and shows a dramatic
increase in the correlation once vaccinations became available to the entire
adult population. By early May, the correlation reached 0.8, a much stronger
relationship than just a month earlier. In the right panel, we see that the
correlation between each state's vaccine uptake time series and its time series
of survey estimates is also quite high, highlighting the survey's ability to
track trends.

There are several interesting points to take away from this. First, the utility
of CTIS data for comparisons between states evidently increased over this time
period---despite Figure 3 of \textcite{Bradley:2021} showing the \textit{ddc}
increasing over the same time period, implying \emph{decreasing} data quality.
Though the \textit{ddc} is intended to serve as an ``index of data quality,''
this suggests that a different framework would be needed to assess the utility
of surveys for tasks other than point estimation.

Second, though the \textit{ddc} decomposition includes a term for ``problem
difficulty'', that term does not quantify the difficulty of \emph{this} problem.
The error decomposition shown in \cref{error-decomposition} is for the error in
estimating a population mean, and the problem difficulty term is \(\sigma_Y\),
the standard deviation of the quantity in the population. But in this problem,
the error \(\bar Y_n - \bar Y_N\) is irrelevant; the goal is to estimate the
\emph{ranking} of states to determine which states have the lowest vaccination
rates. (We could imagine using this, or a similar ranking for vaccine hesitancy,
as part of decisions for prioritizing government resources to encourage
vaccination.) The difficulty of the ranking problem depends on the heterogeneity
of the \emph{states}; if all states have similar vaccination rates, ranking them
well requires very tight margins of error.\footnote{A complete analysis of the
  data's utility for resource allocation would be more complicated: for example,
  if states have very similar vaccination rates, an error in prioritization
  would be less harmful than an error when some states have dramatically lower
  rates than others.}

Let's examine the problem difficulty in more detail, since it will illuminate
the value and weaknesses of the CTIS results.

\subsection{Ranking difficulty and CTIS's biases}

One way to quantify the difficulty of a survey estimation problem is to ask:
What size of random sample would be required to achieve our desired result, on
average? This is the goal of the ``effective sample size'' \(n_\text{eff}\)
calculated by \textcite{Bradley:2021}, which is is ``the size of a simple random
sample that we would expect to exhibit the same level of error as what was
actually observed in a given study'' (page 9). If the level of error observed is
known, for example if we can compare the survey results to a gold standard, the
\textit{ddc} decomposition allows \(n_\text{eff}\) to be estimated.
\textcite{Bradley:2021} calculated that in the case of CTIS, ``a biased national
sample of 250,000 contains no more usable information than a simple random
sample of size 20'' (page 22).

But the estimation problem here is not point estimation. If we are interested in
prioritizing resources between states, the appropriate question is ``What size
of simple random sample would we expect to exhibit the same rank correlation
with the truth as what was actually observed in a given survey?'' We can already
see that \(n_\text{eff} = 20\) is not the answer to this question, because a
sample of size 20 from the United States population cannot provide a ranking of
50 states, let alone one that correlates strongly with the true ranking.

We conducted a simulation study to explore this point further. In each
simulation, we drew a random sample from the United States population, drawing a
simple random sample from each state with size in proportion to its proportion
of the total population. Within each state, we simulated a survey estimate of
the state vaccination rate; this estimate was centered at the CDC-estimated
vaccination rate and drawn from a sampling distribution based on the state
sample size. The resulting estimates were then correlated with the CDC ground
truth data, and this procedure was repeated 1,000 times to yield an average
correlation.

\begin{figure}
  \centering
  \includegraphics[width=0.7\textwidth]{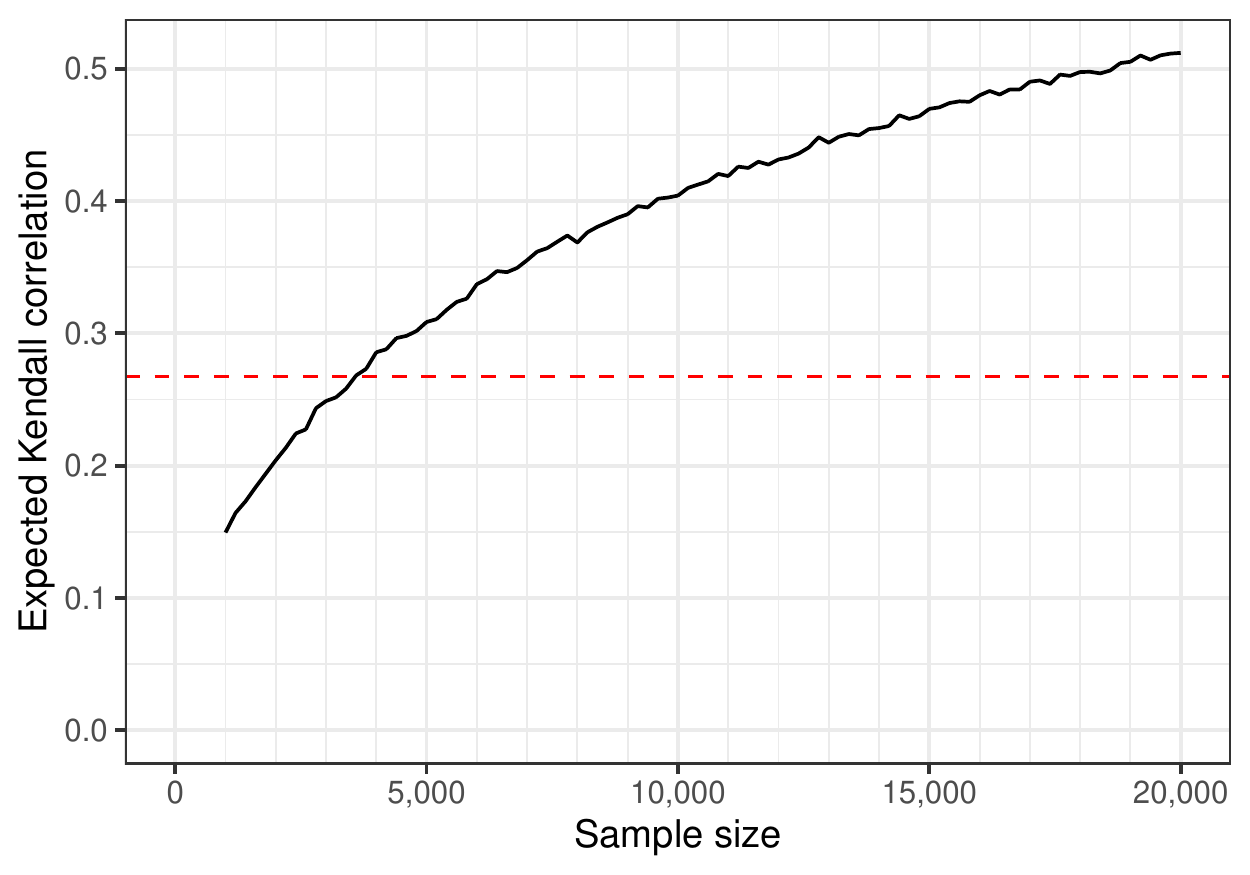}
  \caption{On March 27, 2021, the expected Kendall correlation between
    state-level survey estimates and CDC data, if the state estimates were
    obtained from a random sample of the United States population. CTIS's
    correlation on that date is shown in red.}
  \label{eff-n-march-27}
\end{figure}

\Cref{eff-n-march-27} shows the results of this simulation over a range of
national sample sizes. We can see that on March 27---again, according to
\Cref{cdc-compare-correlation}, the date where CTIS has nearly its worst
correlation---a sample of nearly 4,000 respondents would be required to match
CTIS. And even a sample of \(n=20000\) respondents would, on average, only
barely reach a correlation of 0.5 with the truth, because states had similar
vaccination rates and were difficult to rank.

\begin{figure}
  \centering
  \includegraphics[width=0.7\textwidth]{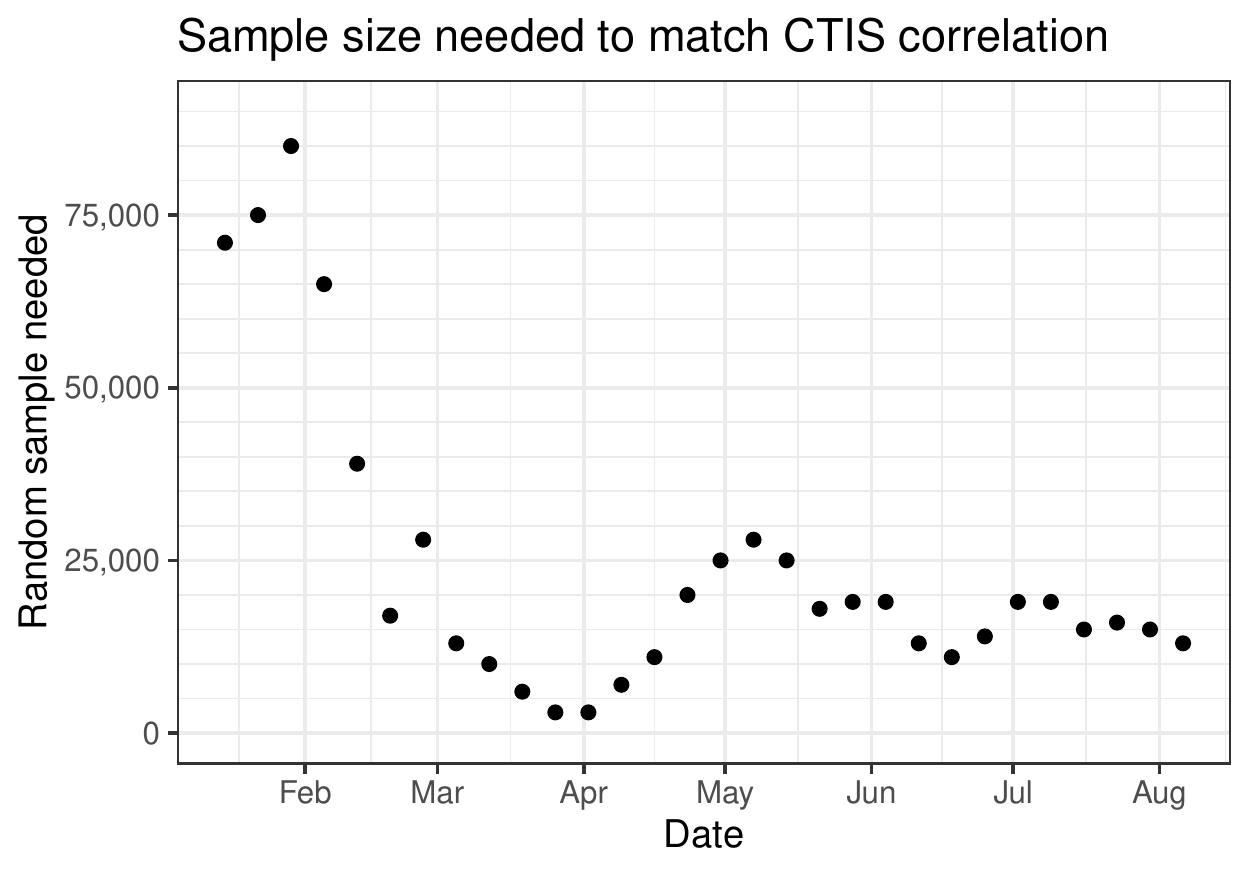}
  \caption{For each week, the random sample size needed to match CTIS's
    Kendall correlation with state-level CDC data.}
  \label{eff-n-time}
\end{figure}

We would expect this to vary over time, since the problem difficulty changed as
state vaccination rates became more heterogeneous. Indeed, \Cref{eff-n-time}
shows how the sample size needed to match CTIS's correlation changed each week,
showing that early on---when all states had very low vaccination rates---a
random sample of size over 70,000 would be required to rank the states as
accurately as CTIS did, while by summer, a sample of size 15--20,000 would be
required instead.

So when \textcite{Bradley:2021} note that large surveys correlate poorly with
ground-truth data, we must first ask ``How hard is this problem?'' We find that
the \textit{ddc} decomposition does not answer this question correctly, and that
the ranking problem was actually quite hard, requiring even an unbiased random
sample to be quite large to perform well. Only with this answer in hand can we
evaluate survey performance. We conclude that while CTIS's sample of size
250,000 does not perform as well as a random sample of the same size would, it
performs \emph{much} better than one would expect from the analysis presented by
\textcite{Bradley:2021}. This implies the spatial variation in sampling bias is
not as severe as claimed, and the claim that ``a biased national sample of
250,000 contains no more usable information than a simple random sample of size
20'' (page 22) does not apply to the ranking task. We believe this claim only
applies to the narrower task of unbiased population point estimation.

Returning to our main argument, we see that \Cref{cdc-compare-time} hints at the
reason for these problems. The \textit{ddc} is defined in terms of the
correlation of two random variables, but the correlation depends on both the
relationship between the variables \emph{and their marginal distributions}.
Vaccination was rapidly increasing over this time period, and so was the
difference between states: in March, the range in true vaccination rates was
limited---between 20 and 40\%---but that range expanded dramatically over six
weeks. This shift in marginal distribution caused both an increase in the
Kendall correlation, and hence the utility of the survey for making comparisons
and allocating resources, and an increase in the \textit{ddc}. We will explore
this problem in more detail in the next section when we explore biases over
time.

\section{Is the bias increasing over time?}
\label{bias-increasing}

Besides making comparisons between states, we have also advocated for CTIS's use
in tracking trends over time. Though we have shown this strategy to be quite
successful for tracking COVID case rates, Figure 3 of \textcite{Bradley:2021}
suggests that both the Census Household Pulse and CTIS surveys experience
decreasing data quality, as measured by the data defect correlation
(\textit{ddc}), over time. Page 12 notes that:
\begin{quote}
  This decomposition suggests that the increasing error in estimates of vaccine
  uptake in Delphi--Facebook and Census Household Pulse is primarily driven by
  increasing \textit{ddc}, which captures the overall impact of the bias in
  coverage, selection, and response.
\end{quote}
They reach this conclusion by comparing CTIS and Household Pulse estimates of
COVID-19 vaccine uptake to CDC data on vaccine distribution, which is reasonably
comprehensive and reliable. Both surveys overestimate vaccination rates, and the
size of the difference increases over time.

But this does not imply that the sampling bias is increasing, as they note:
\begin{quote}
  However, this does not necessarily imply a change in the response mechanism,
  because an identical response mechanism can result in a different \textit{ddc}
  as the correlation between that mechanism and the outcome changes, e.g., an
  individual's vaccination status \(Y\) changes over time.
\end{quote}
We think it is useful to explore the consequences of this caveat in more detail,
since it limits how well the \textit{ddc} can be used as an ``index of data
quality.''

Why could \textit{ddc} be increasing for these surveys? Crucially, the
underlying population quantity---the percentage of American adults who are
vaccinated---is monotonically increasing. For a fixed sampling bias, in a
situation where respondents who are willing to be vaccinated are more likely to
respond to a survey, the estimation bias monotonically increases \emph{even if
  the probability of responding is held fixed for every member of the
  population}. Hence the \textit{ddc} increases even if the data's quality is
fixed. A simple model will help us demonstrate this point.

\subsection{Demonstration}
\label{ddc-demo}

Consider a simplified example. Let \(\ind\) be the indicator function, and let
\(V = \ind(\text{person is vaccinated})\). Let \(R = \ind(\text{responds to
  survey})\). Let the population be composed of two groups of people, denoted by
\(G \in \{1, 2\}\). These groups have both different probabilities of responding
and different probabilities of being vaccinated:
\begin{align*}
  \Pr(G = 1) &= \eta & \Pr(V = 1 \mid G = 1) &= \rho(t) & \Pr(R = 1 \mid G = 1)
                                                          &= 0.02\\
  \Pr(G = 2) &= 1 - \eta & \Pr(V = 1 \mid G = 2) &= \rho(t) / b & \Pr(R = 1 \mid
                                                                  G = 2) &= 0.02/\gamma,
\end{align*}
where
\begin{itemize}
\item \(b \geq 1\) is the differential vaccination rate: if \(b = 2\), members
  of group 2 are half as likely to be vaccinated as members of group 1
\item \(\gamma \geq 1\) is the differential response rate: if \(\gamma = 2\),
  members of group 2 are half as likely to respond to the survey as members of
  group 1. In line with typical response rates for online surveys on the
  Facebook platform, 2\% of group 1 responds.
\end{itemize}
Crucially, the vaccination rate of group 1 is \(\rho(t)\), which depends on time
\(t\) as more vaccines become available. Assume that \(R \perp V \mid G\), i.e.\
\(G\) is the only determinant of vaccination or response probability.

If we conduct a survey of this simplified population, what would be the bias in
the survey's estimated vaccination rate, as a function of the true rate? A
survey of this population would estimate the quantity \(\Pr(V \mid R = 1)\), and
the estimation bias is \(\Pr(V \mid R = 1) - \Pr(V)\). We can see that when
\(\rho(t) = 0\), this bias must be zero, since all respondents would say they
are not vaccinated. (We are ignoring measurement error in this example.)

\begin{figure}
  \centering
  \includegraphics[width=0.8\textwidth]{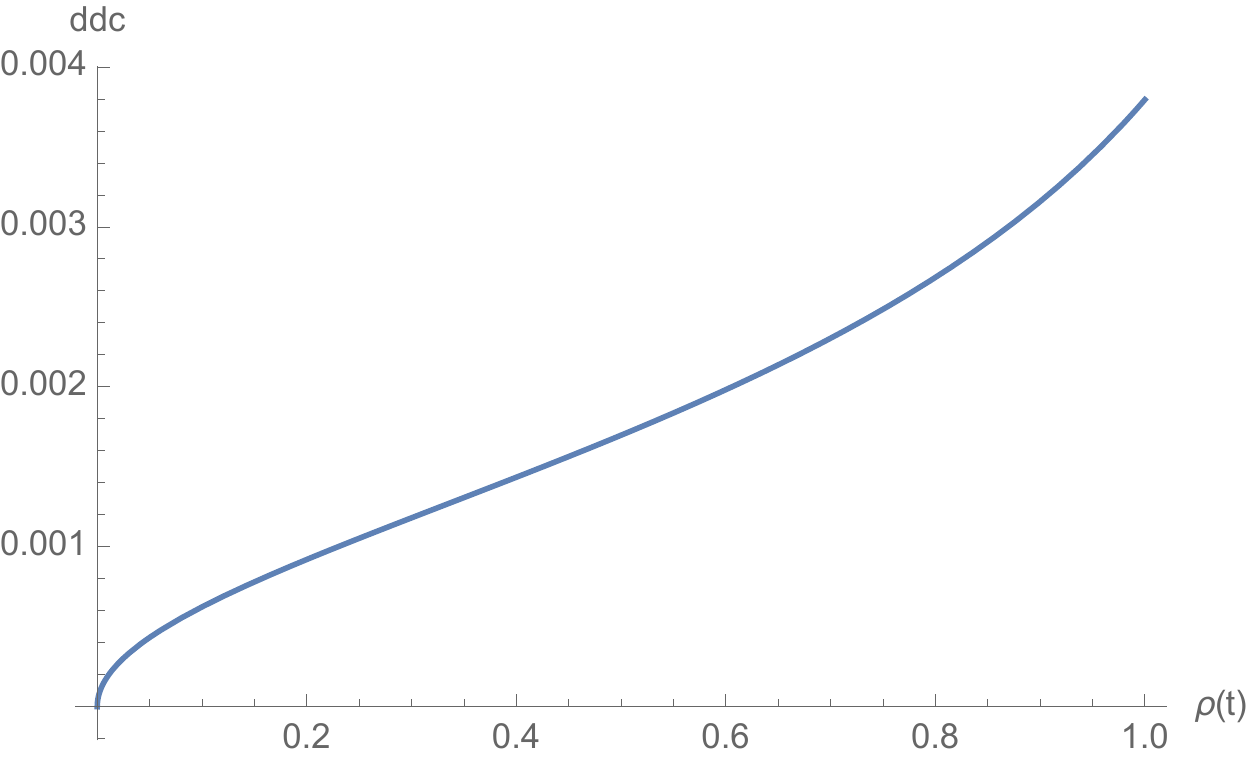}
  \caption{The data defect correlation (\textit{ddc}) as a function of
    \(\rho(t)\), for the simple example in \Cref{ddc-demo}. Here \(b = 2\) and
    \(\eta = 0.5\).}
  \label{ddc-rho}
\end{figure}

We can use Bayes' rule and algebra to work out this bias as a function of
\(\rho(t)\) (see \Cref{appendix-derivation} for details). Assuming a population
of size 250,000,000 and sample of size 30,000 (roughly matching the United
States adult population and a typical daily sample size for CTIS in 2021,
respectively), \Cref{ddc-rho} illustrates how the \textit{ddc} varies with
\(\rho(t)\) when \(b = 2\), \(\gamma = 4\), and \(\eta = 0.5\). (These values
were chosen because one might expect about half the adult population to harbor
views skeptical of vaccines.) This demonstrates that, with no change in response
probabilities, the \textit{ddc} can increase for a survey where vaccine uptake
is associated with response, even if the mechanism for that association is
fixed. Notably, the scale of \textit{ddc} values shown in \Cref{ddc-rho} is
similar to the scale of those in Figure 3 of \textcite{Bradley:2021},
illustrating that a moderate but fixed sampling bias is consistent with the
changing \textit{ddc} presented there.

\subsection{Analysis in the observed data}

\begin{figure}
  \centering
  \includegraphics[width=0.9\textwidth]{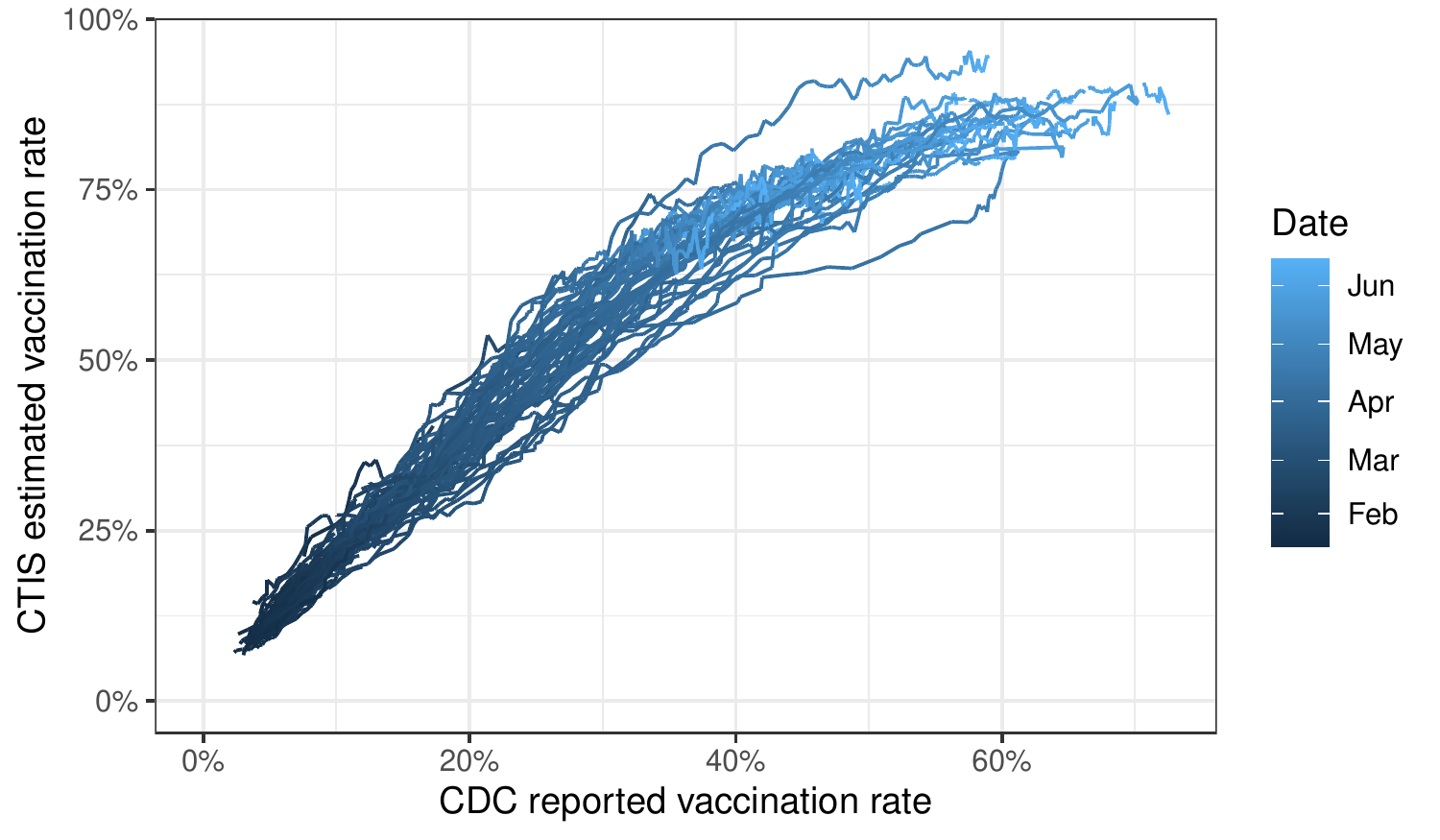}
  \caption{Official CDC estimates of population vaccination rates at each date
    and CTIS estimates of adult vaccination rate on the same date. Each line
    represents one US state or territory over time.}
  \label{vaccination-rate-compare}
\end{figure}

The model in \cref{ddc-demo} is simplified: it does not account for any form of
measurement error in the survey questions (which we will discuss in
\Cref{ddc-survey-limitations}). But it still provides an interesting point of
comparison. The model predicts a linear relationship, with slope greater than 1,
between the vaccination rate and the survey estimates (see
\Cref{appendix-derivation}), and we can look at this relationship in the
observed data easily. \Cref{vaccination-rate-compare} shows the results.

For the first several months of the vaccination campaign, the relationship
appears quite linear, with different slopes for each state. This matches what we
could expect in the simplified model if the response mechanism was fixed in time
but varied between states: the slope is a function of \(\gamma\), \(\eta\), and
\(b\), which we might reasonably expect to vary by state. Hence this does not
indicate the presence of biases that change over time.


Only later, in April and May, do the slopes begin to decrease, reducing the gap
between survey estimates and CDC data. This could reflect shifts in the response
mechanism that \emph{reduce} the bias; for example, perhaps some hesitant
respondents became more willing to both become vaccinated and respond to the
survey. This may also reflect COVID vaccines being made available to children 12
years and older beginning in early May; CTIS only samples respondents 18 and
older and hence cannot track vaccination in this group, though CDC data includes
all vaccinations. The slope change appears to correspond to the time period for
which \textcite[Fig. 3]{Bradley:2021} showed CTIS's \textit{ddc} decreasing.




\subsection{Implications}

First, these results mean that the \textit{ddc} should not be interpreted as
representing the survey bias or data quality per se. We can see that by
remembering that the \textit{ddc} is a correlation between a random variable
\(V\) and each population member's survey response indicator \(R\):
\[
  \ddc = \rho_{V,R} = \corr(V, R).
\]
Just like any correlation, if \(V\) is concentrated around one value, it will be
smaller than if \(V\) has a larger range; that is to say, the \textit{ddc}
depends on both the sampling bias and the marginal distribution of \(V\), and
comparisons of \textit{ddc} between time points or populations with different
distributions of \(V\) are comparisons of \emph{both} sampling bias and \(V\).

This suggests that one cannot compare surveys in different locations or time
periods, or measuring different quantities, by comparing their \textit{ddc}
values; or at least that such comparisons would not reveal which survey is more
methodologically sound. It also suggests that there are other explanations for
the \textit{ddc}'s changes over time than the one presented on page 49:
\begin{quote}
  Delphi-Facebook's \textit{ddc} is higher overall, and shows a stark divergence
  between the two age groups after March 2021. The \textit{ddc} for seniors
  flattens and starts to decrease after an early March peak, whereas the error
  rate for younger adults continues to increase through the month of March 2021,
  and peaks in mid-April, around the time at which all US adults became
  eligible.

  This is consistent with the hypothesis that barriers to vaccine and online
  survey access may be driving some of the observed selection bias in
  Delphi-Facebook. Early in the year, vaccine demand far exceeded supply, and
  there were considerable barriers to access even for eligible adults, e.g.,
  complicated online sign-up processes, ID requirements, and confusion about
  cost.
\end{quote}
This is also consistent with the observation that vaccination rates among
seniors increased rapidly through March, but that the increase slowed in March
as most willing seniors were already vaccinated, causing the \textit{ddc} to
stop rising. Meanwhile, vaccination rates among younger adults rose quickly for
a month or two longer as eligibility expanded. (The drops in \textit{ddc} at the
end of the time period are not well-explained by either hypothesis.) It is not
clear to us how we could distinguish between these two hypotheses using the
survey data alone.

Finally, the relatively consistent bias shown in \Cref{vaccination-rate-compare}
suggests that the survey bias changed only slowly over time. This matches the
survey's design goal of tracking trends over time and detecting sudden changes.
The bias does differ by state (or, more likely and more precisely, by local
demographic factors), explaining why the state-level survey estimates do not
perfectly correlate with CDC data.

\section{Limitations of the data defect framework for surveys}
\label{ddc-survey-limitations}

Above, we have argued that the data defect correlation measures only one aspect
of a dataset's suitability for purpose, that it is difficult to interpret when
population quantities are changing, and that the COVID-19 Trends and Impact
Survey has proven suitable for several important purposes not considered by
\textcite{Bradley:2021}. These are important points to consider, but there lurks
a more fundamental problem: the error decomposition given by
\textcite{Meng:2018} does not directly apply to surveys subject to measurement
error.

\subsection{Interpretation of the ddc with measurement error}

\textcite{Bradley:2021} estimated the \textit{ddc} using \cref{ddc-estimator} by
substituting CDC data, assumed to be reliable, for \(\bar Y_N\) and substituting
survey estimates for \(\bar Y_n\). But a survey does not obtain \(Y_i\); it asks
real humans for \(Y_i\), and real humans may interpret the question differently
than intended, misremember events, feel pressured to give particular answers, or
deliberately lie \parencite{Tourangeau:2000}. Let \(Y_i^*\) denote the response
of subject \(i\) to a survey question asking them about \(Y_i\). The
relationship of \(Y_i^*\) to \(Y_i\) may be complex and depend on
individual-level features, and the survey estimator is the sample mean \(\bar
Y_n^*\), not \(\bar Y_n\).

As \textcite[page 691]{Meng:2018} noted, the error decomposition does not apply
to the error \(\bar Y^*_n - \bar Y_N\) if there is measurement error:
\begin{quote}
  Statistically, [the decomposition] is applicable whenever the recorded values
  of [\(Y\)] can be trusted; for example, if a response is to vote for Clinton,
  it means that the respondent is sufficiently inclined to vote for Clinton at
  the time of response, not anything else. Otherwise we will be dealing with a
  much harder problem of \textit{response bias}, which would require strong
  substantive knowledge and model assumptions [see, e.g., Liu et al.\ (2013)].
  See Shirani-Mehr et al.\ (2018) for a discussion of other types of response
  bias that contribute to the so-called \textit{Total Error} of survey
  estimates.
\end{quote}
\textcite[page 10]{Bradley:2021} acknowledge this point, but state that in this
setting, the data defect correlation ``becomes a more general index of data
quality directly related to classical design effects,'' justifying its use to
evaluate the surveys. However, we do not believe this is true when systematic
measurement error is present. We must examine the derivation in their
Supplementary Material B.1 to spot the limitation.

Suppose we use the sample mean of respondent reports \(Y_i^*\) to estimate
\(\bar Y_n\); denote this sample mean by \(\bar Y_n^*\). The total survey error
is hence \(\bar Y_n^* - \bar Y_N\). If we plug this into \cref{ddc-estimator}
and rearrange, we obtain
\begin{equation}
  \label{ddc-design-effect}
  \hat \rho_{R,Y} \sqrt{N} = \frac{\bar Y_n^* - \bar Y_N}{\sqrt{(1 - f)
      \sigma^2_Y / n}}.
\end{equation}
We denote the quantity in \cref{ddc-design-effect} as \(Z\); \textcite[page
39]{Bradley:2021} argue that ``the expectation of \(Z^2\) with respect to \(R\)
(if it is random) is simply the well-known \textit{design effect}.'' The design
effect is simply the variance of an estimator (under whatever sampling strategy
is used) divided by the variance of a simple random sample estimator applied to
the same population:
\begin{align*}
  \E[Z^2] &= \E\left[\frac{(\bar Y_n^* - \bar Y_N)^2}{(1 - f) \sigma_Y^2/n}\right]\\
          &= \frac{n}{(1 - f)\sigma_Y^2} \E\left[(\bar Y_n^* - \bar Y_N)^2 \right].
\end{align*}
The first term is the reciprocal of the simple random sample variance when
sampling from a finite population; the second term is the variance of \(\bar
Y_n\) \textbf{if and only if} \(\E[\bar Y_n^*] = \bar Y_N\). Hence \(Z^2\) is
related to the design effect if and only if our estimator suffers no systematic
measurement error. In a real survey instrument, this is unlikely.

One can also see this problem through two trivial examples:
\begin{enumerate}
\item Suppose 50\% of the population is vaccinated, a simple random sample is
  taken, and all sampled individuals respond via in-person visits from an
  interviewer. But the interviewer is extremely intimidating, making 100\% of
  respondents report that they are vaccinated. The design effect will be zero
  (because, regardless of the sample, the sample mean will always be 100\%),
  \(\corr(Y, R)\) will also be nearly zero (because it is a simple random
  sample without nonresponse bias), but \(\hat \rho_{R,Y}\) will be large when
  estimated by \cref{ddc-estimator}.

\item Suppose 50\% of the population is vaccinated and the survey is voluntary,
  but it is only completed by vaccinated individuals, and the question text is
  confusing and causes half of respondents to interpret it backwards. The sample
  estimate will hence be 50\% on average. The design effect will be nearly one
  (since we are really estimating what proportion of vaccinated people misread
  the question, and the population mean of that quantity is the same as the
  population vaccination rate) and \(\hat \rho_{R,Y}\) will be nearly zero, even
  though \(\corr(Y, R)\) is large.
\end{enumerate}
As a result, if we plug the total error from a real survey into
\cref{ddc-estimator} and obtain \(\hat \rho_{R,Y}\), the result is neither
\(\corr(R, Y)\) nor is it connected to the survey design effect.
\textcite{Bradley:2021} conducted scenario analyses to study the effects of
survey biases on estimates of vaccine hesitancy (their Section 6), but these
scenario analyses relied on the assumption that \(\hat \rho_{R,Y} \approx
\corr(R, Y)\) (see their Appendix E.1). Because of the limitations above, this
assumption need not be true, though it is unclear to us how this will affect the
results of their scenario analysis; further analytical work and simulations may
be needed to determine if the scenario analysis approach is viable.

In sum, the problem we discussed in \Cref{bias-increasing} is magnified:
increasing \textit{ddc} does not imply increasing sample bias, nor does it
necessarily imply increasing correlation between the sampling mechanism and the
population quantity. It only implies increasing total error, which could come
from several sources, and a small total error could still mask serious sampling
or measurement problems.

\subsection{The total survey error framework}

We have seen that researchers considering using survey data to answer their
research questions must carefully consider whether the data is appropriate, and
that the \textit{ddc} neither answers that question directly (except for the
very specific goal of population point estimation) nor helps researchers
identify the types of error in a survey that might affect their work.
Fortunately, survey methodologists have developed more complete frameworks for
describing sources of errors in surveys, along with methods for evaluating how
each might affect a survey.

Researchers using survey data can apply the Total Survey Error (TSE) framework
\parencite{Biemer:2003,Biemer:2010} in order to understand potential sources of
error and how they could affect the research questions at hand. The TSE
framework breaks down possible sources of error into representation errors
(including sampling errors) and measurement errors. Measurement errors include
problems of validity (when the question text measures a different concept than
was intended by the researcher), measurement error (when errors occur in data
collection due to problems such as inaccurate translations or misunderstanding
by respondents), and processing error (such as data cleaning and coding errors).

In the case of COVID-19 vaccination rates, processing error could affect both
survey estimates and benchmark estimates from official sources, due to
differences across jurisdictions in reporting structures and timelines. But
measurement error in particular could play an important role in the differences
observed in CTIS and Census Household Pulse on the one hand and Axios-Ipsos on
the other.

For example, consider the question text used by each instrument to ascertain the
respondent's vaccination status:
\begin{description}
\item[Axios-Ipsos] Do you personally know anyone who has already received the
  COVID-19 vaccine?
  \begin{itemize}
  \item Yes, I have received the vaccine
  \item Yes, a member of my immediate family
  \item Yes, someone else
  \item No
  \end{itemize}

\item[CTIS] Have you had a COVID-19 vaccination?
  \begin{itemize}
  \item Yes
  \item No
  \item I don't know
  \end{itemize}

\item[Census Household Pulse] Have you received a COVID-19
  vaccine?\footnote{Beginning in Phase 3.2, on July 21, 2021, this text changed
    to ``Have you received at least one dose of a COVID-19 vaccine?}
  \begin{itemize}
  \item Yes
  \item No
  \end{itemize}
\end{description}
Notice that CTIS and Census Household Pulse have very similar items---the CTIS
item was designed to be similar to Household Pulse so it could be interpreted
alongside other Household Pulse data---while the Axios-Ipsos item is somewhat
different. The key question is whether these questions are interchangeable for
the purpose of estimating vaccine uptake, or whether some respondents might
answer these questions differently. Unfortunately that is a question that can
only be answered empirically, not simply from mathematical theory, and we do not
have a direct comparison of the different versions on survey instruments given
to the same population.

One could speculate, however, that social desirability bias is mitigated by the
Axios-Ipsos version of the question. Social desirability bias can occur whenever
respondents feel pressure to give socially acceptable answers, and often occurs
on surveys of sensitive topics. The amount of social desirability bias can be
affected by small changes in wording that make respondents feel more or less
comfortable answering \parencite{Krumpal:2013}. For example, perhaps respondents
feel uncomfortable admitting they are unvaccinated, but are more comfortable if
they can say they at least know someone \emph{else} who is vaccinated. Another
possible explanation is related to the phenomenon of acquiescence, in which
survey respondents preferentially give positive or agreeable answers
\parencite{Krosnick:1999}; the Axios-Ipsos wording, by giving respondents the
opportunity to say ``Yes'' without saying they have personally been vaccinated,
would reduce the effect of acquiescence on estimates of vaccination rate.

We do \emph{not} mean to suggest these are the right explanations for the
difference between survey estimates; they are merely two explanations of many
possible explanations. These hypotheses are at least consistent with the
observed data, which shows CTIS and Census Household Pulse's estimates matching
each other more closely than they match Axios-Ipsos, but there could be many
other explanations, including in their sampling frames, their use of incentives
\parencite{Singer:2013}, their nonresponse biases, the position of the questions
relative to other questions, or their weighting procedures
\parencite{Groves:2010}. The data defect correlation, oblivious as it is to
sources of measurement and specification error, would not help us assess these
possibilities. Additional detailed empirical study is required, possibly
including experiments using different question text or different sampling
strategies.

\section{Discussion}

A natural conclusion of the argument made by \textcite{Bradley:2021} might be
that users of large-scale survey data subject to sampling biases should
carefully examine its suitability for their research goals, for example by
applying the total survey error framework \parencite[chapter 2]{Biemer:2003},
rather than blindly assuming that a large sample size guarantees suitability. We
would endorse this conclusion, which is why it is surprising that
\textcite{Bradley:2021} do not reach it. In fact, they cite a prominent example
of this caution, noting that ``Delphi--Facebook is a widely-scrutinized survey
that, to date, has been used in 10 peer-reviewed publications, including on
important topics of public policy such as mitigation strategies within schools''
(page 22). This is a reference to \textcite{Lessler:2021cu}, who used CTIS data
to track the proportion of children attending in-person schooling over time, and
the mitigation measures applied by each school district. But they did not use
the data unaware of its limitations, as one might assume based on the brief
description given; instead, in preceding work, \textcite{LuptonSmith:2021}
carefully compared schooling data from CTIS to two other schooling datasets that
relied on direct data collection from individual school districts, and hence
provided benchmarks. Their caution was rewarded: they found good consistency
between the datasets and carefully examined limitations in the survey data,
strengthening their subsequent research.

We believe that large and nontraditional datasets, such as administrative
records and massive surveys, have an increasing role to play in modern
science---but that this role must be balanced with a keen awareness of their
limitations. While we believe that \textcite{Bradley:2021} have made important
points deserving of consideration by anyone using a large dataset, we also
believe they have not gone far enough. Because of the limitations of the data
defect correlation framework, it does not provide researchers adequate tools to
evaluate large survey datasets in the context of their specific research goals.
General comments about big data's failings do not help researchers fix them, and
in the face of such warnings, those with an important question and relevant big
data will still be tempted to use the data. We must provide them tools such as
the data defect correlation, but also other tools to estimate measurement error,
measure sampling bias, improve weighting, estimate the effects of potential
biases, and so on. As we have shown here, a single metric (like the
\textit{ddc}) cannot do this alone.

We welcome further work on tools and frameworks to study and characterize large
datasets, and we encourage the broader statistics and survey science communities
to work together to address these challenges. There is a need for additional
tools to understand all components of total survey error
\parencite{Groves:2010}. We believe that a key lesson of ongoing debates in the
statistics community---over reproducibility, hypothesis testing, \(p\) values,
and so on---is that scientific practice is improved by engaging constructively
with scientists to equip them with appropriate tools, rather than through
post-hoc criticism of their results.

\appendix

\section{Derivation of bias example}
\label{appendix-derivation}

In this appendix, we derive some of the results underlying the simple model in
\Cref{ddc-demo}. In that model, we have that the true population vaccination
rate is:
\begin{align*}
  \Pr(V) &= \Pr(V \mid G = 1) \Pr(G = 1) + \Pr(V \mid G = 2) \Pr(G = 2)\\
         &= \eta \rho(t) + (1 - \eta) \rho (t) / b\\
         &= \frac{\rho(t) + (b - 1) \eta \rho(t)}{b}.
\end{align*}
We now work out the conditional probability of vaccination among those who
respond to the survey, which is what the sample proportion among survey
respondents would estimate:
\begin{align*}
  \Pr(V \mid R = 1)
  &= \sum_{g \in \{1,2\}} \Pr(V \mid R = 1, G = g) \Pr(G = g \mid R = 1)\\
  &= \sum_{g \in \{1, 2\}} \frac{\Pr(V \mid G = g) \Pr(R = 1 \mid G = g) \Pr(G =
    g)}{\Pr(R = 1)}\\
  &= \frac{\rho(t) + (b \gamma - 1) \eta \rho(t)}{b + (\gamma - 1) b \eta}.
\end{align*}
The error in the survey estimate is the difference between the vaccination rate
among survey respondents and the vaccination rate in the entire population:
\begin{equation}\label{expected-total-error}
  \Pr(V \mid R = 1) - \Pr(V) = \frac{(b - 1)(\gamma - 1)(\eta - 1) \eta}{b +
    (\gamma - 1) b \eta} \rho(t).
\end{equation}
The bias \(\Pr(V \mid R = 1) - \Pr(V)\) is hence an increasing function of
\(\rho(t)\) whose slope depends on \(b\) (and goes to zero when \(b = 1\)).

If we obtain a particular sample of \(n\) respondents from the population and
let \(Y_i \in \{0, 1\}\) indicate whether each person is vaccinated, the sample
mean \(\bar Y_n\) is an estimator of \(\Pr(V \mid R = 1)\). \Cref{ddc-rho} was
produced by plugging \cref{expected-total-error} into \cref{ddc-estimator}, the
\textit{ddc} estimator, in place of the error \(\bar Y_n - \bar Y_N\).

\printbibliography

\end{document}